\documentclass[11pt]{article}

\usepackage{cite}
\usepackage{amssymb}
\usepackage{amsmath}
\usepackage{bm}

\begin{document}

\title{The Algebraic Way.}
\author{B. J. Hiley\footnote{E-mail address b.hiley@bbk.ac.uk.}.}
\date{Physics Department, UCL and TPRU, Birkbeck, University of London, Malet Street, London WC1E 7HX.\\ \vspace{0.4cm}(22 Janurary 2015) }
\maketitle

\begin{abstract}
In this paper we examine in detail the non-commutative symplectic algebra underlying quantum dynamics. We show that this algebra contains both the Weyl-von Neumann algebra and the Moyal algebra.  The latter contains the Wigner distribution as the kernel of the density matrix. The underlying non-commutative geometry can be projected into either of two Abelian spaces, so-called `shadow phase spaces'.  One of these is  the phase space of Bohmian mechanics, showing that it is a fragment of the basic underlying algebra.  The algebraic approach is much richer, giving rise to two fundamental dynamical time development equations which reduce to the Liouville equation and the Hamilton-Jacobi equation in the classical limit. They also include the Schr\"{o}dinger equation and its wave function, showing that these features are a partial aspect of the more general non-commutative structure.  We discuss briefly the properties of this more general mathematical background from which the non-commutative symplectic algebra emerges.

\end{abstract}

\section{Introduction}

\begin{quote}
The basic principle of the algebraic approach is to avoid starting with a specific Hilbert space scheme and rather to emphasise that the {\em primary objects } of the theory are the fields (or the observables) considered as purely {\em algebraic quantities}, together with their linear combinations, products and limits in the appropriate topology (Emch~\cite{ge72}).
\end{quote}

In order to understand the motivation for ``The Algebraic Way" we need to recall the origins of quantum theory.  History tells us that the first pioneering papers to develop a mathematical approach to quantum phenomena were those of Born, Dirac, Heisenberg and Jordan~\cite{mbpj25, mbwhpj26, pd25}. 
Their attempts to accommodate  the Ritz-Rydberg combination principle, an empirical discovery in atomic spectra, into a dynamical theory forced the classical variables of  Hamiltonian dynamics to  be replaced by non-commuting analogues.  

With the emerging law of non-commutative multiplication,  the need for a matrix representation of $x$ and $p$ was soon recognised even though the physical meaning of such a change was unclear.  To the physicist, these matrix representations opened up a new field of unfamiliar non-commutative algebras with which they were not very comfortable and since the mathematics itself appeared to have no obvious physical interpretation, the approach was eventually abandoned in favour of the Schr\"{o}dinger wave mechanics approach.

This approach gained greater impetus when particles were found to exhibit the wave-like behaviour predicted by de Broglie.  These experimental results encouraged Schr\"{o}dinger~\cite{es26} to look for what he
 called a ``Hamiltonian undulatory mechanics" by modifying the Hamilton-Jacobi offshoot of Hamiltonian dynamics.  His motivation came from noting that while ray optics could be explained using equations that were analogous in form to Hamilton's equation of motion for particles, the Hamilton-Jacobi theory contained surfaces of constant action, which suggested an analogy with the wave fronts used in the Huygens construction to explain interference phenomena in light.  
 
 This exploration led Schr\"{o}dinger~\cite{es26} to a differential equation which immediately produced energy levels that conformed with the Ritz-Rydberg data.  The mathematical techniques involved in solving differential equations were well known to physicists at that time and the faith in this equation was further reinforced with the introduction of Born's probability postulate, establishing the relation between wave and particle.  Although this relation was not entirely clear conceptually,  it enabled the formalism to be applied with outstanding success.

Conceptually the wave and algebraic approaches were very different, one being based on a very familiar wave phenomenon, the other being based on an unfamiliar non-commutative dynamics with no obvious interpretation.  Soon Schr\"{o}dinger himself showed how the two approaches were related and, since the techniques for solving the Schr\"{o}dinger equation were very familiar, this approach became established as {\em the} way to understand the physics of quantum phenomena.  

Nevertheless many conceptual problems remained, generating many different interpretations, some naive others quite bizarre, all based on the assumption that the Schr\"{o}dinger equation tells the {\em whole} story, not only for understanding  individual experimental phenomena, but in defining what ultimately constitutes `reality'.  However as we will show in this paper that,  in spite of its great successes, it is only a  part of the whole story.  In order to see this, we need to return to examine the details of the original algebraic approach in some detail.

\section{Representations}

Before discussing these issues, I would like to briefly highlight the relevant  features of the Schr\"{o}dinger approach that we will need in order to motivate our presentation.  Of course, we will start with the Schr\"{o}dinger equation, even though it is not clear exactly how it was derived from the Hamilton-Jacobi theory:
\begin{eqnarray*}
i\hbar \frac{\partial \psi}{\partial t}=\hat H\psi\quad \mbox{with}\quad H(x,p)\rightarrow\hat H(\hat X,\hat P)
\end{eqnarray*}
where the classical Hamiltonian $H(x,p)$ is replaced by its operator form $\hat H(\hat X,\hat P)$.  

To work with the equation, we must go to a specific representation.  It customary to use the Schr\"{o}dinger representation for which 
\begin{eqnarray*}
\hat X\rightarrow x\quad\hat P\rightarrow -i\hbar \frac{\partial}{\partial x}\quad\psi\rightarrow\psi(x,t)
\end{eqnarray*}
so that we are working in configuration space $(x_1,x_2,\dots,x_n)$.

However this is not the only representation.  We can use the $p-$representation where
\begin{eqnarray*}
\hat X\rightarrow i\hbar \frac{\partial}{\partial p}\quad\hat P \rightarrow p\quad\psi\rightarrow\psi(p,t)
\end{eqnarray*}
so that in this case we are working in momentum space $(p_1,p_2,\dots  p_n)$.  Again we have the oscillator representation where
\begin{eqnarray*}
\hat X\rightarrow (a^\dag +a)/\sqrt{2}\quad \hat P\rightarrow i(a^\dag -a)/\sqrt{2}\quad \hat N=a^\dag a.
\end{eqnarray*}
This representation enables us to work more easily with an arbitrary number of particles and is essential for quantum field theory.  Of course the Schr\"{o}dinger representation is favoured because we believe that quantum processes actually occur in Minkowski space-time.

Although there is an abundance of mathematical representations, the Stone-von Neumann theorem proves that all irreducible representations  are unitarily equivalent.  By this we mean that if there are two unitary representations, $\pi_1$ and $\pi_2$, in their respective Hilbert spaces ${\cal H}_1$ and ${\cal H}_2$,
\begin{eqnarray*}
\pi_1:G\rightarrow U({\cal {H}}_1)\quad\mbox{and}\quad \pi_2:G\rightarrow U({\cal{H}}_2)
\end{eqnarray*}
and there exists an operator $A:{\cal {H}}_1\rightarrow {\cal {H}}_2$, then these representations are equivalent iff there exists an operator $A$ such that
\begin{eqnarray*}
A\pi_1(g)=\pi_2(g)A\quad\forall g\in G.
\end{eqnarray*}
Having established mathematical equivalence, we are left with the question, ``Are the representations also physically equivalent?"  This, in turn, leaves another question ``Of what mathematical structures are they representations?"

\section{Common Symmetries underlying both Classical and Quantum Mechanics}

It is generally believed that quantum phenomena ``demand a fundamental modification of the basic physical concepts and laws"~\cite{llel77}.  In 
other words we require a totally different description from that used in classical physics.  However there are some obvious similarities in the {\em form} of the dynamical equations of motion. 
In classical mechanics, Hamilton's equations of motion can be written in the form
\begin{eqnarray}
\dot x_i=\{x_i,H\};\quad \dot p_i=\{p_i,H\}\quad\mbox{and generally}\quad \dot f(x_i,p_i)=\{f(x_i,p_i),H\}	\label{eq:ham}
\end{eqnarray}
where $H$ is the Hamiltonian and  $\{.,.\}$ are the Poisson brackets defined by
\begin{eqnarray*}
\{f,g\}=\sum_i\left[\frac{\partial f}{\partial x_i}\frac{\partial g}{\partial p_i}-\frac{\partial f}{\partial p_i}\frac{\partial g}{\partial x_i}	\right],
\end{eqnarray*}
giving the special case $\{x_i,p_j\}=\delta _{ij}$.

On the other hand, in quantum mechanics, Heisenberg's equations of motion appear in the form
\begin{eqnarray}
i\hbar\frac{ d\hat X_i}{dt}=[\hat X_i,\hat H];\quad i\hbar\frac{ d\hat P_i}{dt}=[\hat P_i,\hat H]\quad \mbox{and generally}\quad i\hbar \frac{d\hat F}{dt}=[\hat F,\hat H].
\label{eq:heis}
\end{eqnarray}
Here $\hat H$ is the classical Hamiltonian where $x$ and $p$ are replaced by operators $\hat X$ and $\hat P$ and $[.,.]$ are the commutator brackets defined by
\begin{eqnarray*}
[\hat F,\hat G]=\hat F\hat G-\hat G\hat F
\end{eqnarray*}
giving in the special case $[\hat X_i,\hat P_j]=i\hbar\delta_{ij}$.  

The similarity in the {\em form} of the two sets of equations is quite remarkable, in spite of the differences in the nature of the elements involved.  The classical  equations of motion are ordinary functions on a continuous phase space while, in the quantum case, they are operators acting on vectors in an abstract Hilbert space.  However there is one other striking similarity.  They are both invariant under the Heisenberg group.

The Heisenberg equations of motion directly use elements of the Heisenberg (Lie) algebra defined by the canonical commutation relations 
\begin{eqnarray*}
[\hat X_i,\hat X_j]=[\hat P_i,\hat P_j]=0,\quad [\hat X_i,\hat P_j]=\delta_{ij}\hat T,\quad [\hat X_i,\hat T]=[\hat P_i,\hat T]=0.
\end{eqnarray*}
Here we have written $\hat T=i\hbar \hat I$ for convenience, so that the elements $(\hat X_i, \hat P_i,\hat T)$ generate  the Heisenberg group,  $H_n$.
 
On the other hand, the classical dynamical variables are {\em representations} of the Heisenberg algebra in which commutators are replaced by Poisson brackets. Thus the Heisenberg group is not only significant in the quantum domain but also operates in the classical domain.  In fact it plays a vital role in radar theory~\cite{lafg95}, which is in no way a quantum phenomenon.

There is a further invariance which is more directly seen in the classical mechanics in the dynamical equations of motion (\ref{eq:ham}).  They are are invariant under transformations of the symplectic group $Sp(2n)$ (i.e. canonical transformations) for a $2n$-dimensional phase space.  These transformations leave invariant the antisymmetric bilinear form $\omega(x,p;x',p')=xp'-x'p$.  Although one can prove this directly, it can also be thought of as arising from the group of automorphisms of the underlying Heisenberg group.

If we write two elements of the Heisenberg group in the form
\begin{eqnarray*}
\hat U=\sum_{i=1}^n x_i\hat X_i+p_i\hat P_i +t\hat T,\quad\hat U'=\sum_{i=1}^n x'_i\hat X_i+p'_i\hat P_i +t'\hat T,
\end{eqnarray*}
we find
\begin{eqnarray}
[\hat U,\hat U']=\omega(x,p;x',p')\hat T, \label{eq:hs}
\end{eqnarray}
where $\omega$ is again an antisymmetric bilinear form.  The appearance of $\omega(x,p;x',p')$ in equation (\ref{eq:hs}) implies that the Heisenberg group and, hence the Heisenberg equations of motion, are invariant under the group of symplectic transformations.  In other words the group of automorphisms of the Heisenberg group is the symplectic group.

This means that the mathematical structure underlying both classical and quantum dynamical behaviour arises from symplectic geometry. It turns out that, in the quantum case, the symplectic geometry is  non-commutative, while in the classical case, it is commutative.  Although these structures are clearly related mathematically, we still have a puzzle as to why there is no trace of an underlying phase space in the quantum algebra and, even if we were to find one, then how would it accommodate the Heisenberg uncertainty principle? 

\section{How do we Relate the Quantum Algebra to the Phase Space Description?}

To find the role of phase space in quantum mechanics, we must put aside any objections based on the uncertainty principle and follow some early work of von Neumann~\cite{vn31}.
Let us consider, not only translations in $x$-space, but also translations in $p$-space.  As is well known, space translations can be described using the Taylor expansion so that
\begin{eqnarray*}
f(x+a)=\exp\left[\alpha\frac{\partial}{\partial x}\right]f(x).\quad\quad\alpha\in \mathbb R
\end{eqnarray*}
In the case of a translation in momentum space, we may similarly write
\begin{eqnarray*}
g(p+\beta)=\exp\left[\beta\frac{\partial}{\partial p}\right]g(p).\quad\quad\beta\in \mathbb R
\end{eqnarray*}
By recalling the Schr\"{o}dinger representation, we can write these exponentials in operator form, namely
\begin{eqnarray*}
\hat U(\alpha)=\exp(i\alpha \hat P)\quad\mbox{and}\quad\hat V(\beta)=\exp(i\beta\hat X).
\end{eqnarray*}
We immediately see that these translations do not commute because
\begin{eqnarray*}
\hat U(\alpha)\hat V(\beta)=e^{i\alpha\beta}\hat V(\beta)\hat U(\alpha).
\end{eqnarray*}
The operators $\hat U(\alpha)$ and $\hat V(\beta)$ generate the Weyl-von Neumann algebra.

\subsection{Relation between the Weyl-von Neumann Algebra and Hilbert Space}

To make the link with the Hilbert space formalism, von Neumann introduced the algebraic element
\begin{eqnarray}
\hat S(\alpha,\beta)=e^{-i\alpha\beta/2}\hat U(\alpha)\hat V(\beta).	\label{eq:sab}
\end{eqnarray}
Then for a system described by $|\psi\rangle$ in the usual Hilbert space, we can form the expectation value
\begin{eqnarray*}
S_\psi(\alpha,\beta)=\langle \psi|\hat S(\alpha,\beta)|\psi\rangle.
\end{eqnarray*}
 von Neumann then shows that {\em any} linear operator $\hat A$ can be symbolically written as
\begin{eqnarray*}
\hat A=\int\int a(\alpha,\beta)\hat S(\alpha, \beta)d\alpha d\beta.
\end{eqnarray*}
This leads to a  quantum expectation value of the operator $\hat A$, via
\begin{eqnarray*}
\langle\psi|\hat A|\psi\rangle=\int\int a(\alpha, \beta)S_\psi(\alpha,\beta)d\alpha d\beta
\end{eqnarray*}
where the kernel $a(\alpha,\beta)$ is defined by
\begin{eqnarray*}
a(\alpha,\beta)=\int\langle \alpha+\gamma|\hat A|\alpha-\gamma\rangle e^{-2i\beta\gamma}d\gamma.
\end{eqnarray*}
In this way we can completely reproduce the expectation values of quantum mechanics in terms of functions of real variables $(\alpha, \beta)$.  I refer the reader to von Neumann for the details.

In the Weyl-von Neumann approach, then, the operators of the quantum formalism are replaced by differential functions on the $\alpha,\beta$-space.  However von Neumann made no attempt to explain the physical meaning of the space spanned  by the parameters $\alpha$ and $\beta$.  Nevertheless one fact emerges: the multiplication of two of these functions, say, $a(\alpha,\beta)$ and $b(\alpha,\beta)$, must be non-commutative in order to reproduce the results of quantum mechanics.

Suppose $\hat A\leftrightarrow a(\alpha,\beta)$ and $\hat B\leftrightarrow b(\alpha,\beta)$, then if $\hat A.\hat B\leftrightarrow a(\alpha,\beta)\star b(\alpha,\beta)$, von Neumann shows that
\begin{eqnarray}
a(\alpha,\beta)\star b(\alpha,\beta)=\int\int e^{[i(\alpha\beta'-\alpha'\beta)/2]}
a(\alpha-\alpha',\beta-\beta')b(\alpha',\beta')d\alpha' d\beta'. \label{eq:starp}
\end{eqnarray}
Not only is this star product\footnote{This product, although first defined by von Neumann, is now known as the Moyal star product.} non-commutative, it is also non-local.  Thus non-locality appears as a basic feature of the $\alpha,\beta$ plane, so if we want to replace the operators of the quantum formalism by continuous functions, then the resulting structure must be non-local.

\subsection{Moyal's Contribution to the Physical Meaning of the Weyl-von Neumann Algebra}

Moyal~\cite{jm49} arrived at exactly the same mathematical structure as 
von Neumann but by starting from a very different approach. He was trying to understand the nature of the statistics that is needed in quantum mechanics, so he asked the question ``How can we generalise the statistics of random variables if these variables are non-commutative?"

With a pair of commutative random variables $X, Y$, one defines the expectation values by introducing the characteristic function $e^{i(Xt+Ys)}$. Then the expectation value of some function $f_{X,Y}(x,y)$ is
\begin{eqnarray*}
\phi_{X,Y}(t,s)=E\left[e^{i(Xt+Ys)}\right]=\int\int e^{i(xt+ys)}f_{X,Y}(x,y)dxdy.
\end{eqnarray*}
Moyal proposed that, in the non-commutative case, the characteristic function be replaced by $\langle\psi|e^{i(\alpha\hat P+\beta\hat X)}|\psi\rangle$ so that we can form the function
\begin{eqnarray}
F_\psi(x,p)=\frac{1}{4\pi}\int\int \langle\psi|e^{i(\alpha \hat P+\beta\hat X)}|\psi\rangle e^{i(\alpha p+\beta x)}d\alpha d\beta.	\label{eq:dist}
\end{eqnarray}
He then proposed that the average of any quantum operator $\hat A$ can be found using
\begin{eqnarray}
\langle\psi|\hat A|\psi\rangle=\int\int a(x,p)F_\psi(x,p)dxdp.	\label{eq:cexpt}
\end{eqnarray}
Note that Moyal has now introduced the two parameters $x$ and $p$ through the Fourier transform (\ref{eq:dist}) and since we are dealing with a single particle, it has been assumed that these parameters are the position and momentum of a single particle.  If that were where the case, then from the form of equation (\ref{eq:cexpt}), we could regard $F_\psi(x,p)$ as a probability distribution for the particle having coordinates $(x,p)$ and we can then regard equation (\ref{eq:cexpt}) as giving the quantum expectation value for the operator $\hat A$ by averaging  $a(x,p)$ over a phase space.  

\noindent There are two difficulties in making such an assumption.  
\begin{enumerate}
\item 
As is well known, $F_\psi(x,p)$ is the Wigner function\footnote{We will show this later in section 4.3.} and can become negative.  The assumption that $F_\psi(x,p)$ is a probability density then opens up a debate as to the validity of the whole approach.  However we will show that $F_\psi(x,p)$ is not a probability distribution, but the kernel of a density matrix which is not necessarily  positive definite or even real.  Thus it is the {\em interpretation} of $F_\psi(x,p)$ being  a probability distribution that is not valid, not the method in which it arises, so we can follow Feynman~\cite{rf87} and use equation (\ref{eq:cexpt}) as a valid way to evaluate the quantum expectation values without worrying about the appearance of negative values of $F_{\psi}(x,p)$.  We need to remember that we are dealing with a non-commutative structure and not simply averaging over classical coordinates.
\item
As is not so well known, the parameters $(x,p)$ are not the position and momentum of a localised particle, but the mean values of a cell in phase space associated with the particle. Thus in this approach, the particle cannot be considered as a point-like object.  Rather it is a non-local distribution of energy, the quantum blob~\cite{mdg12,bh03}.  This region, which we associate with the particle, explains the non-local nature of the $\star$ product.
\end{enumerate}

\subsection{Relation to the Wigner Distribution}

We will now show the function $F_\psi(x,p)$ is, in fact, the one particle Wigner function, the many-body generalisation of which was first introduced by Wigner~\cite{ew32} to discuss the thermodynamic properties of quantum systems.

 First consider the operator 
$\hat S(\alpha,\beta)$ defined in equation (\ref{eq:sab}) written in a slightly modified form
\begin{eqnarray*}
\hat S'(\alpha,\beta):=e^{i\alpha\beta}\hat U(\alpha)\hat V(\beta)=e^{i\alpha\hat P/2}e^{i\beta\hat X}e^{i\alpha\hat P/2}.
\end{eqnarray*}
 It is not difficult to show that
\begin{eqnarray*}
\langle\psi|\hat S'(\alpha,\beta)|\psi\rangle=\int\psi^*(x-\alpha/2)e^{i\beta x}\psi(x+\alpha/2)dx.
\end{eqnarray*}
By taking the Fourier transform, we find
\begin{eqnarray}
F_\psi(x,p)=\frac{1}{2\pi}\int\psi^*(x-\alpha/2)e^{-i\alpha p}\psi(x+\alpha/2)d\alpha,	\label{eq:Wfn}
\end{eqnarray}
which we recognise as the Wigner function. Thus we see the Wigner function is intimately connected with the Weyl-von Neumann-Moyal algebraic approach.

\subsection{Non-Commutative Phase Space}

In order to confirm that we are dealing with a non-commutative phase space, we will follow Moyal~\cite{jm49}, who showed that the star-product (\ref{eq:starp}) can be written in a more convenient way,
\begin{eqnarray}
a(x,p)\star b(x,p)=a(x,p)\exp[i\hbar(\overleftarrow \partial_x\overrightarrow\partial_p-\overrightarrow\partial_x\overleftarrow\partial_p)/2]b(x,p).\label{eq:mprod}
\end{eqnarray}
It is not difficult to show that this expression when applied to $x$ and $p$ gives\footnote{In the earlier sections we have used the parameter $p$ without giving it a physical meaning. If we want to interpret it as a momentum, we must replace it by $p/\hbar$.}
\begin{eqnarray*}
x\star p-p\star x=i\hbar.
\end{eqnarray*}
Thus we see that although we are dealing with functions of ordinary real $(x,p)$  variables, the usual commutative inner product must be replaced by a non-commutative product.  

Once we have a non-commutative product we must distinguish between left and right multiplication.  However  we find it easier to take this into account by introducing two types of bracket, namely,
\begin{eqnarray*}
\{a,b\}_{MB}=\frac{a\star b- b\star a}{i\hbar}\quad \mbox{and}\quad \{a,b\}_{BB}=\frac{a\star b+b\star a}{2}.
\end{eqnarray*}
The first is the Moyal bracket, while the second is the Baker bracket (or the Jordan product).  Using the expression for the product (\ref{eq:mprod}), it is easy to show
\begin{eqnarray*}
\{a,b\}_{MB}=a(x,p)\sin[\hbar(\overleftarrow \partial_x\overrightarrow\partial_p-\overrightarrow\partial_x\overleftarrow\partial_p)/2]b(x,p)
\end{eqnarray*}
and
\begin{eqnarray*}
 \{a,b\}_{BB}=a(x,p)\cos[\hbar(\overleftarrow \partial_x\overrightarrow\partial_p-\overrightarrow\partial_x\overleftarrow\partial_p)/2]b(x,p).
\end{eqnarray*}
The importance of these brackets is that they become classical objects in the  limit  $O(\hbar^2)$.  The Moyal bracket becomes the Poisson bracket 
\begin{eqnarray*}
\{a,b\}_{MB}=\{a,b\}_{PB}+O(\hbar^2)=[\partial_xa\partial_pb-\partial_pa\partial_xb]+O(\hbar^2)
\end{eqnarray*}
while the Baker bracket to the same approximation reduces to the simple product
\begin{eqnarray*}
 \{a,b\}_{BB}=ab+O(\hbar^2).
\end{eqnarray*}
Thus we see that the non-local $\star$-product now becomes the local inner product used in classical mechanics.  Thus in one single  formalism we have a way of dealing with both quantum and classical mechanics\footnote{ These results form the basis of deformation quantisation~\cite{ahph02}.}.
\section{Non-Commutative Dynamics: the Phase Space Approach}

As we have seen, an important lesson when dealing with a non-commutative algebra is to carefully distinguish between left and right multiplication\footnote{More formally the mathematical structure of quantum mechanics is a bimodule.}. We have been able to avoid this distinction by going to
 the Schr\"{o}dinger representation which gives a simpler algorithm that only uses left multiplication.  To exploit the full implications of the non-commutative structure we have to go deeper.
 
To define the dynamics in such a mathematical structure,  we have to consider the following two equations
\begin{eqnarray}
H(x,p)\star F_\psi(x,p,t)=i(2\pi)^{-1}\int e^{-i\tau p}\psi^*(x-\tau/2,t)\overrightarrow\partial _t\psi(x+\tau/2,t)d\tau	\label{eq:ltrans}
\end{eqnarray}
and
\begin{eqnarray}
F_\psi(x,p,t)\star H(x,p)=-i(2\pi)^{-1}\int e^{-i\tau p}\psi^*(x-\tau/2,t)\overleftarrow\partial _t\psi(x+\tau/2,t)d\tau.	\label{eq:rtrans}
\end{eqnarray}
Subtracting these two equations gives us one time development equation expressed in terms of the Moyal bracket:
\begin{eqnarray}
\partial_tF_\psi=(H\star F_\psi-F_\psi\star H)/2i=\{H,F_\psi\}_{MB}.	\label{eq:mbe}
\end{eqnarray}
While by adding the two equations, we get another time development equation expressed in terms of the Baker bracket~\cite{gb58}:

\begin{eqnarray}
2\{H,F_\psi\}_{BB}
=i(2\pi)^{-i}\int e^{-i\tau p}[\psi^*(x-\tau/2,t)\overleftrightarrow\partial_t\psi(x+\tau/2,t)]d\tau	\label{eq:tbb}
\end{eqnarray}
where
\begin{eqnarray}
\frac{\psi^*\overleftrightarrow\partial_t\psi}{\psi^*\psi}=\frac{[\psi^*\overrightarrow\partial_t\psi -\psi^*\overleftarrow\partial_t\psi ]}{\psi^*\psi}.
\label{eq:janus}
\end{eqnarray}
It should be noted that we need {\em both} equations to get a complete description of quantum mechanics.   For a more detailed discussion see Zachos~\cite{cz02}.

We have already seen that equation (\ref{eq:mbe}) leads to the classical Liouville equation in the classical limit.  To see what equation (\ref{eq:tbb}) gives in the classical limit, let us put $\psi=Re^{iS}$ into equation (\ref{eq:tbb}), expand out and then take the limit to $O(\hbar^2)$.  We find 
\begin{eqnarray*}
\{H,F_\psi\}_{BB}=H.F_\psi+O(\hbar^2)=-2(\partial_tS)F_\psi +O(\hbar^2)
\end{eqnarray*}
which then gives the classical Hamilton-Jacobi equation,
\begin{eqnarray*}
\frac{\partial S}{\partial t}+H=0.
\end{eqnarray*}
A related approach to the classical limit will be found in Schleich~\cite{ws05}.  

This is a very interesting result when we recall that Schr\"{o}dinger actually started from the classical Hamilton-Jacobi equation in order to find an ``Hamiltonian undulatory mechanics".  One of the reasons why he was forced to guess his equation was because he not did fully appreciate the significance of non-commutativity.

\section{Where does the Bohm Approach fit in to this Structure?}	

\subsection{Conditional Expectation Values in the Moyal Approach}\label{subsec:ba}
Since the Moyal algebra gives the correct quantum expectation values of quantum operators by averaging over a symplectic phase space and since the Bohm approach gives the same expectation values using what seems to be a different phase space defined in terms of $(x, p=\nabla S)$, there surely must be a relation between these two approaches.  To bring out this relationship, let us follow Moyal and treat $F_\psi(x,p)$ as a quasi-probability distribution.  We can then define the conditional expectation value of the momentum.  

A value of this momentum can be obtained from the general relation given by Moyal~\cite{jm49}, namely
\begin{eqnarray}
\rho(x)\overline {p^n}=\int p^n F_\psi(x,p)dp=\left(\frac{\hbar}{2i}\right)^n[(\partial_{x_1}-\partial_{x_2})\psi(x_1)\psi(x_2)]_{x_{1}=x_{2}=x}.
\label{eq:condx}
\end{eqnarray}
For $n=1$ we find, by writing $\psi=Re^{iS}$, that
\begin{eqnarray*}
\overline p(x)=\frac{1}{2i}[\psi^*\nabla\psi-(\nabla\psi^*)\psi]=\nabla S(x).
\end{eqnarray*}
This is identical to the Bohm momentum defined by the relation $p=\nabla S$, the so called ``guidance relation".  However in the approach we are exploring here, there are no waves of any form and the notion of guiding wave is meaningless.  Everything that emerges is a consequence of the non-commutative symplectic geometry.

This connection between the Bohm momentum and the conditional expectation value of the momentum can be made even stronger.  Moyal shows that by starting from the Heisenberg equations of motion, the transport of the momentum $\overline p(x,t)$ is given by
\begin{eqnarray*}
\partial_t(\rho \overline p_k)+\sum_i \partial_{x_i}\overline{(\rho p_k\partial_{x_i}H)}+\rho\partial_{x_k}\overline H=0.
\end{eqnarray*}
Then after some work and again writing $\psi=Re^{iS}$, Moyal finds
\begin{eqnarray*}
\frac{\partial}{\partial x_k}\left[\frac{\partial S}{\partial t}+\overline H-\frac{\nabla^2 \rho}{8m\rho}\right]=0.
\end{eqnarray*}
If we choose $\overline H=\overline {p^2}/2m+V$ where
\begin{eqnarray*}
\overline {p^2}=(\nabla S)^2-\frac{\hbar^2}{2}\left(\frac{\nabla R}{R}\right)^2+\frac{\hbar^2}{4}\frac{\nabla^2 \rho}{\rho}.
\end{eqnarray*}
Then
\begin{eqnarray}
\frac{\partial S}{\partial t}+\overline H-\frac{\nabla^2 \rho}{8m\rho}=\frac{\partial S}{\partial t}+\frac{1}{2m}(\nabla S)^2+V-\frac{1}{2m}\frac{\nabla^2 R}{R}=0.	\label{eq:qhj}
\end{eqnarray}
Here the RHS of equation (\ref{eq:qhj}) is the quantum Hamilton-Jacobi equation, the real part of the Schr\"{o}dinger equation that plays a key role  in the Bohm approach~\cite{dbbh93}.  But since the Moyal algebra contains the Bohm approach, and in fact is exactly the von Neumann algebra (i.e. an algebra upon which quantum mechanics is based) then clearly the Moyal and the Bohm approach are simply different aspects of precisely the same mathematical structure.

Full details of the above derivations can be found in the appendix of the original Moyal paper~\cite{jm49}.  Further details of the relation between the Moyal and the Bohm approach can be found in Hiley~\cite{bh03}.

\subsection{Shadow Manifolds}

What the previous subsection \ref{subsec:ba}  shows is that if we take the variable $x$ as one axis of the phase space, we can take $\overline p$ to be the other axis of the phase space.  Thus we have constructed a phase space out of the variables $(x,\overline p)$.  In this phase space, the time development equation is the quantum Hamilton-Jacobi equation
\begin{eqnarray}
\partial_t S(x,t)+(\nabla_x S(x,t))^2/2m +Q_x(x,t) +V(x,t)=0.	\label{eq:qhjx1}
\end{eqnarray}
Here the quantum potential, $Q_x(x,t)$, is given by
\begin{eqnarray*}
Q_x(x,t)=-\frac{1}{2m}\left(\frac{\nabla^2R(x,t)}{R(x,t)}\right).
\end{eqnarray*}
Thus we can construct trajectories in this $(x,\overline p)$ space.

However notice that the distribution $F_\psi(x,p)$ is symmetric in $x$ and $p$ so that we can also find the conditional expectation value of the position, $\overline x(p,t)$, in terms of the momentum $p$.  We will again follow  Moyal and define this value $\overline x$ as
\begin{eqnarray*}
\rho(p)\overline x=\int xF_\phi(x,p)dx=\int x\psi^*(x')\psi(x'')\delta[x-(x'+x'')/2]e^{ip(x'-x'')}dxdx'dx'',
\end{eqnarray*}
which in the $p$-representation takes the simpler form
\begin{eqnarray*}
\rho(p)\overline x=\int xF_\phi(x,p)dx=\frac{1}{2i}\left[(\partial_{p_1}-\partial_{p_2})\phi^*(p_1)\phi(p_2)\right]_{p=p_1=p_2}.
\end{eqnarray*}
Writing $\phi(p)=R(p)e^{iS(p)}$, we find the conditional expectation value of the position, $\overline x(p)$, given the value of $p$ is
\begin{eqnarray*}
\overline x(p)=-\nabla_p S(p).
\end{eqnarray*}
Again in analogy with the previous case, we  have another quantum Hamilton-Jacobi equation, only this time in $p$-space.  Thus
\begin{eqnarray}
\partial_tS(p,t) + p^{2}/2m + Q_p(p,t)+
V(-\nabla_pS(p,t),t) = 0,	\label{eq:qhjp1}
\end{eqnarray}
where
\begin{eqnarray}
Q_p(p,t)=- \frac{1}{2mR_{p}}\left(\frac{\partial ^{2}R_{p}}{\partial
p^{2}}\right)
\end{eqnarray}
is the quantum potential in a second phase space constructed in terms of
the coordinates  $(\overline x=-\nabla_pS, p)$.  An example of how this works for the case of a particle in a potential $V(x)=Ax^3$ will be found in Brown and Hiley~\cite{mbbh00} where more details of the whole approach are given.

Thus we find that there are, at least, two shadow phase spaces we can access.   Each gives a different phase space picture of the same overall algebraic structure, a feature that has already been recognised in the Wigner approach by Leibfried {\em et al.}~\cite{l98} who call these spaces shadow phase spaces, a term Hiley~\cite{bh04} has also used.

These shadow spaces are an example of what Bohm calls `explicate orders' in his general notion of the implicate order~\cite{db80}. In this case the algebraic structure defines the implicate 
order, while the two shadow phase spaces are a pair of explicate orders.
One should note that both equations  (\ref{eq:mbe}) and (\ref{eq:tbb}) do not contain quantum potentials explicitly.  They only appear explicitly in equations (\ref{eq:qhjx1}) and (\ref{eq:qhjp1}), namely at the level 
 of conditional expectation values.   One should also note that in the classical limit $\overline p\rightarrow p$ and $\overline x\rightarrow x$, so that, in this limit, both quantum potentials vanish and we have one unique phase space.

\section{Non-Commutative Dynamics: the Algebraic Approach}

\subsection{Operator Equations}

We can get more insight into this whole approach by returning to the operator approach and exploiting the one-to-one relation $\hat A\leftrightarrow a(x,p)$.  This means we should be able to form the operator equivalent of the two equations (\ref{eq:mbe}) and (\ref{eq:tbb}).  In order to motivate this, let us return to consider how the Schr\"{o}dinger equation emerges from the Heisenberg equation for the time development of the density operator $\hat \rho$,
\begin{eqnarray}
i\frac{d\hat\rho}{dt}=[\hat H, \hat \rho]	.	\label{eq:hdme}
\end{eqnarray}
Let us follow Dirac~\cite{pd30}
and write $\hat\rho=\hat \psi.\hat \phi$.  Notice that both $\hat \psi$ and $\hat\phi$ are {\em operators}, not vectors in a Hilbert space.  Substituting this expression into equation (\ref{eq:hdme}),  we get
\begin{eqnarray*}
i\frac{d\hat\psi}{dt}\hat \phi+i\psi\frac{d\hat\phi}{dt}=(\hat H\hat\psi)\hat\phi-\hat\psi(\hat\phi \hat H).
\end{eqnarray*}
Notice we can actually form this equation by subtracting the following two Schr\"{o}dinger-like equations
\begin{eqnarray}
i\frac{d\hat\psi}{dt}=\hat H\hat\psi\hspace{5cm}	\label{eq:Ltrans}\\
\mbox{and}\hspace{11cm}\nonumber\\
-i\frac{d\hat\phi}{dt}=\hat\phi\hat H.\hspace{5cm}	\label{eq:Rtrans}
\end{eqnarray}
We say `Schr\"{o}dinger-like' because $\hat \psi$ and $\hat \phi$ are elements of the operator algebra.  Notice the order of the operators in these two equations;  in equation (\ref{eq:Ltrans}) the operators act from the left, while in equation (\ref{eq:Rtrans}) the operators act from the right.  In fact these equations are left and right translation Schr\"{o}dinger equations, the analogues of equations (\ref{eq:ltrans}) and (\ref{eq:rtrans}) proposed in the von Neumann-Moyal algebra.

Recall that to obtain equation (\ref{eq:mbe}), we subtracted equations (\ref{eq:ltrans}) and (\ref{eq:rtrans}), so we see that the Heisenberg equation of motion can be formed by subtracting equations (\ref{eq:Ltrans}) and (\ref{eq:Rtrans}).    There is a clear analogy with the bra and ket vectors, but here $\hat\psi$ and $\hat\phi$ are taken to be elements of the non-commuting algebra, not elements of an external abstract Hilbert space.  $\hat \psi$ and $\hat \phi$  are, in fact, elements of a specific left and right ideal respectively that exists within the non-commuting symplectic algebra itself.  The implications of this for any possible physical interpretation have been discussed in Hiley~\cite{bh01} and Hiley and Callaghan~\cite{bhbc10}.  

Thus in our approach all the elements we use appear in the algebra itself and there is no essential need to introduce an exterior Hilbert space, although this alternative is available if required for ease of calculation.  This then shows clearly that the Schr\"{o}dinger equation is, as Bohr~\cite{nb61} claimed, merely an algorithm for calculating the outcome of given experimental situations.  But unlike Bohr, we are giving attention to the algebra, in this case the non-commutative symplectic group algebra.   It is this algebra that provides a complete mathematical description of the quantum dynamics.  

We will now bring  out this algebraic structure more  clearly by 
adopting a change of notation, in which  `operators' simply become elements of the algebra because they `operate' on themselves.  Thus we will drop  the `hats' and write $\hat \psi\rightarrow \Psi_L$ and $\hat \phi\rightarrow \Phi_R$.  Here $\Psi_L$ is an element of a suitable left ideal and $\Phi_R$ an element of some suitable right ideal defined by the physics of the problem we are considering.  These elements contain all the information about the state of the system.  Mathematically they are central features of the symplectic Clifford algebra~\cite{ac90}.  Similar features appear in the orthogonal Clifford algebra used to describe the spin and relativistic properties of quantum phenomena~\cite{bhbc12}.  A detailed discussion of how one chooses these ideals will be found in that paper.

\subsection{Left/Right Algebraic Equations}

Let us replace the density operator $\hat \rho$ of a pure state by  $\rho=\Psi_L\Psi_R$, where $\Psi_L$ is a left ideal in the algebra and $\Psi_R$ is the right ideal.  Then the left and right equations of motion are 
\begin{eqnarray*}
i\frac{d\Psi_L}{dt}=H\Psi_L\quad \mbox{and}\quad -i\frac{d\Psi_R}{dt}=\Psi_RH.
\end{eqnarray*}
Next we form
\begin{eqnarray*}
i(\overrightarrow \partial_t\Psi_L)\Psi_R=(\overrightarrow H\Psi_L)\Psi_R\quad\mbox{and}\quad-i\Psi_L(\Psi_R\overleftarrow\partial_t)=\Psi_L(\Psi_R\overleftarrow H).
\end{eqnarray*}
Now we can subtract and add these two equations as before and obtain the two algebraic equations
\begin{eqnarray}
i\left[(\overrightarrow\partial_t\Psi_L)\Psi_R+\Psi_L(\Psi_R\overleftarrow\partial_t)\right]=(\overrightarrow H\Psi_L)\Psi_R-\Psi_L(\Psi_R\overleftarrow H)	\label{eq:A}\\
i\left[(\overrightarrow\partial_t\Psi_L)\Psi_R-\Psi_L(\Psi_R\overleftarrow\partial_t)\right]=(\overrightarrow H\Psi_L)\Psi_R+\Psi_L(\Psi_R\overleftarrow H).	\label{eq:B}
\end{eqnarray}
Since we are writing $\rho=\Psi_L\Psi_R$,  equation (\ref{eq:A}) can be written in the form
\begin{eqnarray}
i\partial_t\rho=[H,\rho]_-.	\label{eq:qle}
\end{eqnarray}
This is, in fact, just the quantum Liouville equation.  Equation (\ref{eq:B}) can be written in the form 
\begin{eqnarray}
i\Psi_R\overleftrightarrow\partial_t\Psi_L=[H,\rho]_+	\label{eq:qec}
\end{eqnarray}
where we have used definition (\ref{eq:janus}).
This equation is simply the expression for the conservation of energy. Thus equations (\ref{eq:qle}) and (\ref{eq:qec}) then are the algebraic equivalents of (\ref{eq:mbe}) and (\ref{eq:tbb}) and give a complete algebraic description of a single quantum system.

\subsection{Emergence of the Bohm Approach through Projections}

In the previous sub-section we showed equations (\ref{eq:qle}) and (\ref{eq:qec}) to be the defining equations for the time development of a single quantum system in terms of the non-commutative symplectic structure.  Notice once again that there is no explicit quantum potential in these equations.

To see how these equations are related to the usual Hilbert space approach, we first introduce a projection operator $\Pi_a=|a\rangle\langle a|$ and apply it to each equation in turn.  We obtain
\begin{eqnarray*}
i\frac{\partial P(a)}{\partial t}+\langle[\rho,H]_-\rangle_a=0,\hspace{0.5cm}\\
2P(a)\frac{\partial S}{\partial t}+\langle[\rho,H]_+\rangle_a=0.
\end{eqnarray*}
Here $P(a)$ is the probability of finding the system in the quantum state $\psi(a)$ which we have written in polar form $\psi(a)=R(a)e^{iS(a)}$.  

In order to get a feel for this approach, it is useful to consider particular examples.  Therefore let us consider the harmonic oscillator,  $H=\frac{p^2}{2m}+\frac{Kx^2}{2}$ for its simplicity and for the fact that it  is symmetric in $x$ and $p$. We will choose two specific projection operators,  $\Pi_x=|x\rangle\langle x|$ and 
$\Pi_p=|p\rangle\langle p|$.  

We will begin by projecting into the $x$-representation using $\Pi_x=|x\rangle\langle x|$ to obtain
\begin{eqnarray}
\frac{\partial P(x)}{\partial t}+\nabla_x.\left(P(x)\frac{\nabla_xS_x}{m}\right)=0\hspace{2cm}	\label{eq:lex}\\
\frac{\partial S_x}{\partial t}+\frac{1}{2m}\left(\frac{\partial S_x}{\partial x}\right)^2-\frac{1}{2mR_x}\left(\frac{\partial^2R_x}{\partial x^2}\right)+\frac{Kx^2}{2}=0.	\label{eq:qhjx}
\end{eqnarray}
Thus we see that equation (\ref{eq:lex}) is the Liouville equation which is the expression for the conservation of probability in the $x$-representation.  Equation (\ref{eq:qhjx}) is the quantum Hamilton-Jacobi equation in the $x$-representation that appears in Bohmian mechanics.

Let us now project  into the $p$-representation by choosing the projection operator $\Pi_p=|p\rangle\langle p|$ to obtain
\begin{eqnarray}
\frac{\partial P_p}{\partial t}+\nabla_p.\left(P_p\frac{\nabla_pS_p}{m}\right)=0 \hspace{1.5cm}	\label{eq:lep}\\
\frac{\partial S_p}{\partial t}+\frac{p^2}{2m}-\frac{K}{2R_p}\left(\frac{\partial^2R}{\partial p^2}\right)+\frac{K}{2}\left(\frac{\partial S_p}{\partial p}\right)^2.	\label{eq:qhjp}
\end{eqnarray}
Notice the appearance again of a quantum potential $Q_p=-\frac{K}{2R_p}\left(\frac{\partial^2R}{\partial p^2}\right)$.  Thus we see the quantum potential becomes manifest only as a result of the projections.  Notice that when the quantum potential is negligible, we recover the classical behaviour, equations (\ref{eq:qhjx}) and (\ref{eq:qhjp}) being related by a canonical transformation.
Although we have illustrated these projections for the harmonic oscillator, it follows trivially that they work for any general Hamiltonian.  

Thus projections from the non-commutative algebraic time development  equations (\ref{eq:qle}) and (\ref{eq:qec}) produce exactly the same results as obtained from the two von Neumann-Moyal equations  (\ref{eq:mbe}) and (\ref{eq:tbb}).  Both lead to the same pair of shadow phase spaces.  Both produce the same quantum Hamilton-Jacobi equations, namely, equations (\ref{eq:qhjx1}) and (\ref{eq:qhjp1}). 

\section{Conditional Expectation Values from the Algebra}

Let us now return to our original motivation, namely, that the primary mathematical structures necessary to describe quantum phenomena are non-commutative geometric algebras.  In this paper we have concentrated on the non-commutative symplectic geometry, restricting ourselves to specific examples to motivate the general method. In a series of papers Hiley and Callaghan~\cite{bhbc10, bhbc10a, bhbc12} have shown how the orthogonal Clifford algebras can be used to describe the spin and relativistic properties of quantum systems.  

These two algebraic approaches are very similar in their mathematical structure, so there is clearly a more general structure of which these algebras are specific examples.  Indeed they are both simple examples of von Neumann algebras and general methods for handling these non-commutative algebras now exist~\cite{deyk98}.  

We will be particularly interested in their relevance to non-commutative probability theory, and in particular, the appearance of conditional expectation values in these structures, which has non-commutative integration theory at its heart~\cite{ies53}.
  We have seen the need to consider left and right differentiation, so that the inverse of differentiation, namely, integration has to take this two-sidedness into account.
Equation (\ref{eq:cexpt}) has been interpreted as providing the expectation value of $a(x,p)$ taken over $F_\psi(x,p)$, treating it as if it were a {\em classical} probability density.   When it was subsequently discovered that $F_\psi(x,p)$ can become negative, alarm bells may have sounded as has been discussed in~\cite{mb45,  rf82, rf87} and more recently in~\cite{hh12}.
 Yet in spite of these difficulties, the expectation values $\langle \psi|\hat A|\psi\rangle$ calculated by these methods always turn out to be positive.  
 
The explanation of these results lies in non-commutative measure theory, particularly in the papers of  Umegaki~\cite{hu54} and Jones~\cite{vj83}.  What Umegaki shows is that a positive definite conditional expectation value always exists in a sub-algebra N of a type II factor von Neumann algebra M, which is the type of algebra we are discussing in this paper.  In particular the conditional expectation $E_N:M\rightarrow N$ is defined by the relation $tr(E_Ny)=tr(xy)$ for $x\in M$ and  $y\in N$.
The map $E_N$ is normal and has the following properties:
\begin{eqnarray*}
E_N(axb)=aE_N(x)b\quad\mbox{for}\quad x\in M, a,b\in M
\end{eqnarray*}
\begin{eqnarray*}
E_N(x^*)=E_N(x)^*\quad\forall x\in M
\end{eqnarray*}
\begin{eqnarray*}
E_N(x^*)E_N(x)\leqq E_N(x^*x)\quad \mbox{and}\quad E_N(x^*x)=0\Rightarrow x=0.
\end{eqnarray*}

Since the von Neumann-Moyal algebra we are discussing here is a type II von Neumann algebra, a trace exists and it remains to evaluate this trace for the two possible projections from the $(x,p)$ algebra to the two Abelian sub-algebras, one spanned by $x$ and the other by $p$.  Our case is trivial since we are considering the special case of a single particle.  

One of the projections we have introduced is $E_P: (x,p)\rightarrow (x)$ which was defined by equation (\ref{eq:condx}).  A careful examination of the origins of $F_\psi(x,p)$ shows that it is actually the kernel of the density matrix itself.  This result has already been pointed out in Hiley~\cite{bh04}
but we will outline the argument briefly again here.

Let us start with the density operator $\hat \rho_\psi=|\psi\rangle \langle \psi|$ and form $\rho_\psi(x_1,x_2)=\psi^*(x_1)\psi(x_2)$ which is the kernel of the density matrix~\cite{mdg06}.  Now let us go to the momentum representation and write
\begin{eqnarray*}
\psi(x)=\frac{1}{\sqrt{(2\pi)}}\int\phi(p)e^{ipx}dp.
\end{eqnarray*}
Then the density kernel can be written as
\begin{eqnarray*}
\rho_\psi(x_1,x_2)=\frac{1}{2\pi}\int\int\phi^*(p_1)e^{-ix_1p_1}\phi(p_2)e^{ix_2p_2}dp_1dp_2.
\end{eqnarray*}
Now let us change co-ordinates to
\begin{eqnarray*}
X=(x_1+x_2)/2\quad\eta=x_2-x_1\quad\mbox{and}\quad P=(p_1+p_2)/2\quad\pi=p_2-p_1,
\end{eqnarray*}
so that the density kernel can be written in the form
\begin{eqnarray*}
\rho_\psi(X,\eta)=\frac{1}{2\pi}\int\int\phi^*(P-\pi/2)e^{iX\pi}\phi(P+\pi/2)e^{i\eta P}dPd\pi.
\end{eqnarray*}
Take the Fourier transform
\begin{eqnarray*}
\rho_\psi(X,\eta)=\int F_\psi(X,P)e^{i\eta P}dP
\end{eqnarray*}
and find
\begin{eqnarray}
F_\psi(X,P)=\frac{1}{2\pi}\int\phi^*(P-\pi/2)e^{iX\pi}\phi(P+\pi/2)d\pi.	\label{eq:ptox}
\end{eqnarray}
Recalling that
\begin{eqnarray*}
\phi^*(P-\pi/2)=\frac{1}{\sqrt{2\pi}}\int\psi^*(x_1)e^{-i(P-\pi/2)x_1}dx_1
\end{eqnarray*}
\begin{eqnarray*}
\phi^*(P+\pi/2)=\frac{1}{\sqrt{2\pi}}\int\psi^*(x_2)e^{-i(P+\pi/2)x_2}dx_2.
\end{eqnarray*}
Using these in equation (\ref{eq:ptox}), we find
\begin{eqnarray*}
F_\psi(X,P)=\frac{1}{2\pi}\int\psi^*(X-\eta/2)e^{-i\eta P}\psi(X+\eta/2)d\eta
\end{eqnarray*}
which is just the expression we used in equation (\ref{eq:Wfn}) with $\eta=\alpha$.

Notice in this construction that the resulting Wigner function is a function in the $(X,P)$ phase space.  This phase space has been constructed from a pair of points in $(x_1,x_2)$ configuration space  and the coordinates $(X,P)$ are the mean position and mean momentum of a cell in an $(x,p)$ phase space.  Thus the Wigner function $F_\psi(X,P)$ is a density matrix over a cell constructed in the underlying $(x,p)$ classical phase space.  
We have kept our arguments deliberately simple to arrive at this result.  A rigorous geometric approach that produces this result and its generalisation can be found in Cari\~{n}ena {\em et al}~\cite{jcjc99}.

The first point to notice is that the Wigner function is a complex density matrix, not a probability density.  This shows why it is incorrect to regard $F_\psi(x,p)$ as a probability distribution of particle positions and momenta. Thus the worries about negative and  complex ``probabilities" are totally unfounded~\cite{hh12}.  

The second point to notice is that the Wigner approach, when applied to a single particle, is non-local depending on a region rather than a single point.  This means we must represent the particle by a region in phase space, namely,   the ``quantum blob" introduced by de Gosson~\cite{mdg12}.  However this non-locality should not be surprising because as we have already pointed out, the $\star$-product is non-local.  The fact that non-locality is an essential feature of the description should again not be surprising.  Indeed the phase space must be non-local otherwise we would be in violation of the uncertainty principle.  That the $\star$-product must be a non-local product has already been pointed out by Gracia-Bondia and V\'arilly~\cite{jgbjv87, jgbjv87a}.  Indeed further details of the mathematical structure lying behind some of the results discussed in this paper will be found in these papers. 

\section{Conclusion}

The aim of this paper has been to show that the algebraic structure of the quantum operators defined by von Neumann~\cite{vn31} and later developed by Moyal~\cite{jm49} gives a more general mathematical structure in which the usual Schr\"{o}dinger representation with its wave function provide but a partial mathematical account of quantum phenomena.  Elsewhere~\cite{h11} we have shown that the information contained in the wave function can be encoded in the algebra in terms of certain ideals already contained in the algebra itself.  Hence there is no fundamental need to postulate an external Hilbert space,  and this is in accord with the principle outlined in the above quotation taken from Emch~\cite{ge72}, namely, that the primary objects of the quantum formalism should be purely algebraic quantities.

The geometries underlying these structures are non-commutative in general and by concentrating on a non-commutative symplectic geometry,  we have shown that the quantum dynamics can be described either by the elements of an abstract algebra or by functions on a generalised phase space.  The multiplication rule for combining these functions is necessarily the non-commutative $\star$-product introduced by von Neumann~\cite{vn31} and Moyal~\cite{jm49}.

Moyal's contribution was to show how the algebra generalised classical statistics to a non-commutative statistics that emerges from a more general non-commutative probability theory~\cite{deyk98}.  By recognising this generalisation, we have shown that the Wigner function emerges from a  representation of the kernel of the density matrix.  We argue that it is therefore  incorrect to regard this kernel as a probability density.  Furthermore this fact explains why the negative values of the Wigner function present no difficulty.

 Within this theory we can introduce conditional expectation values from which Bohmian mechanics emerges under the assumption that space-time is basic. But one has an $x\leftrightarrow p$ symmetry  in the algebra so that it is possible it define an alternative ``mechanics" taking the momentum space as basic.  Thus the Bohm approach does contain the $(x,p)$ symmetry that Heisenberg claimed it lacked~\cite{wh58}.
Moreover this symmetry produces shadow phase spaces as used in ~\cite{l98}.  In Bohm's implicate order, these are what he calls explicate orders.  We have also shown how these shadow manifolds merge into a single commutative phase space in the classical limit.

We noted that the $\star$-product is a non-local product, as does~\cite{jgbjv87}.  Furthermore we have shown that the kernel of the density matrix describes a cell-like structure, rather than a point particle in phase space.  Again this suggests that the quantum particle is represented by a region of the underlying non-commutative symplectic space, so that the quantum formalism is basically non-local in a radically new way even for the single particle, locality arising only at the classical limit. 

\section{Acknowledgements}

I would like to thank Robert Callaghan, Maurice de Gosson, Glen Dennis and David Robson for their invaluable and enthusiastic discussions.



\bibliography{myfile}{}
\bibliographystyle{plain}

\end{document}